\documentclass[%
 reprint,
 amsmath,amssymb,
 aps,
pra,
]{revtex4-2}

\usepackage{graphicx}
\usepackage{dcolumn}
\usepackage{bm}
\usepackage{xcolor}

\begin{document}

\preprint{APS/123-QED}

\title{Broadband Heterodyne Microwave Detection using Rydberg Atoms with High Sensitivity}

\author{Hsuan-Jui Su, Shao-Cheng Fang, Ting-An Li, Chen-Hao Chang, Yu-Chi Chen, and Yi-Hsin Chen}%
 \email{yihsin.chen@mail.nsysu.edu.tw}
\affiliation{%
Department of Physics, National Sun Yat-sen University, Kaohsiung 80424, Taiwan
}%

\date{\today}
\newcommand{\FigOne}{
\begin{figure}[t]
\centering
\includegraphics[width=1 \linewidth]{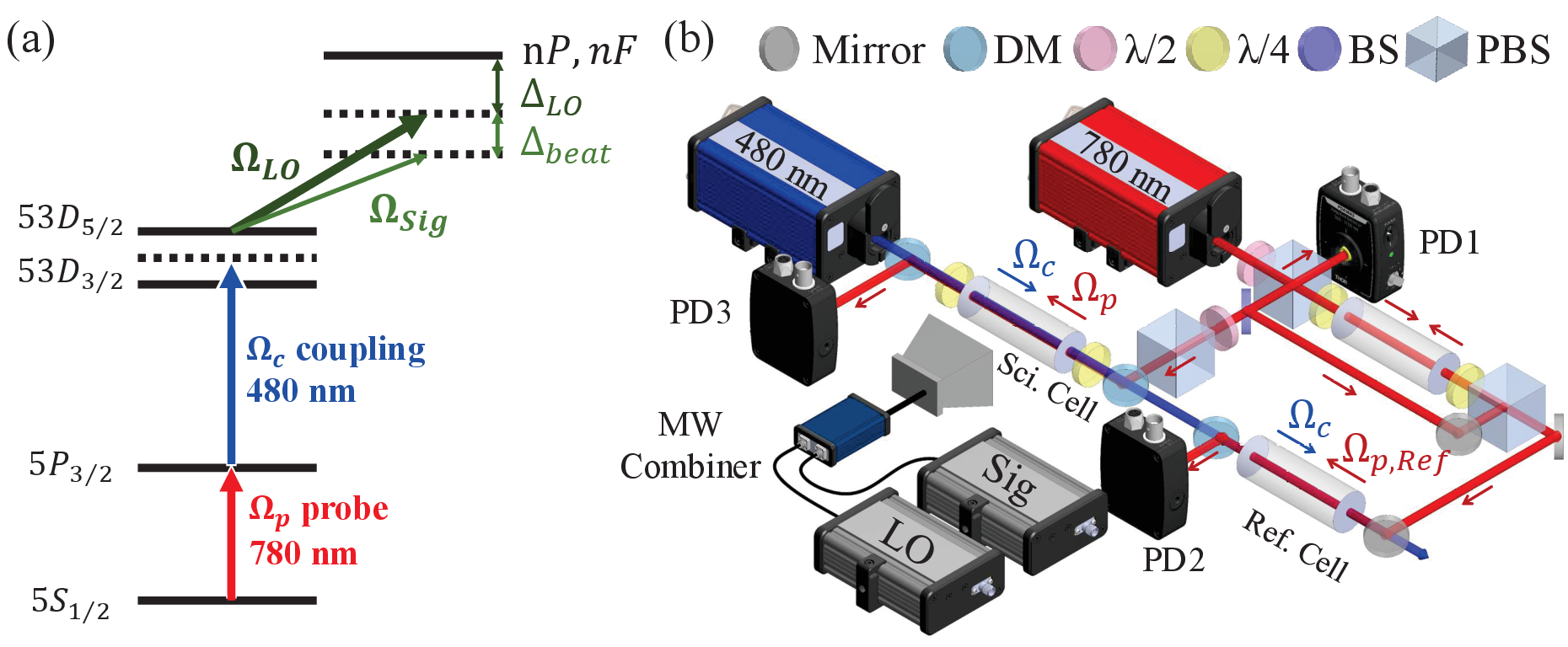}
\caption{(a) Energy level diagram for $^{87}\rm{Rb}$. Atoms are excited to the $|53D\rangle$ Rydberg states from the ground state $|5S_{1/2}\rangle$ via the intermediate state $|5P_{3/2}\rangle$. Two MW fields with different powers and frequencies are applied, which drive the transition to the target states $|54P_{3/2}\rangle$ (14.23 GHz) and $|52F\rangle$ (15.59 GHz). 
The stronger microwave field is designated as the LO field, while the weaker one serves as the signal field for measuring the minimum detectable electric field. The LO frequency is tuned to the transition between the $|53D_{5/2}\rangle$ state and either the $|54P_{3/2}\rangle$ or $|52F\rangle$ state, with a detuning of $\Delta_{LO}$. The signal field frequency is then offset from the LO frequency by a beat frequency difference of $\Delta_{beat}$.
(b) Schematic of the experimental setup. The probe field is frequency-locked using saturation absorption spectroscopy, while the coupling field scans across the $|53D\rangle$ states. The MW fields are generated by two separate signal generators, combined using a power combiner, and emitted from a horn antenna. A separate reference cell, denoted as the Ref. Cell, is used for laser frequency calibration.}
\label{fig:setup}
\end{figure}
}

\newcommand{\FigTwo}{
\begin{figure}[t]
\centering
\includegraphics[width=0.9 \linewidth]{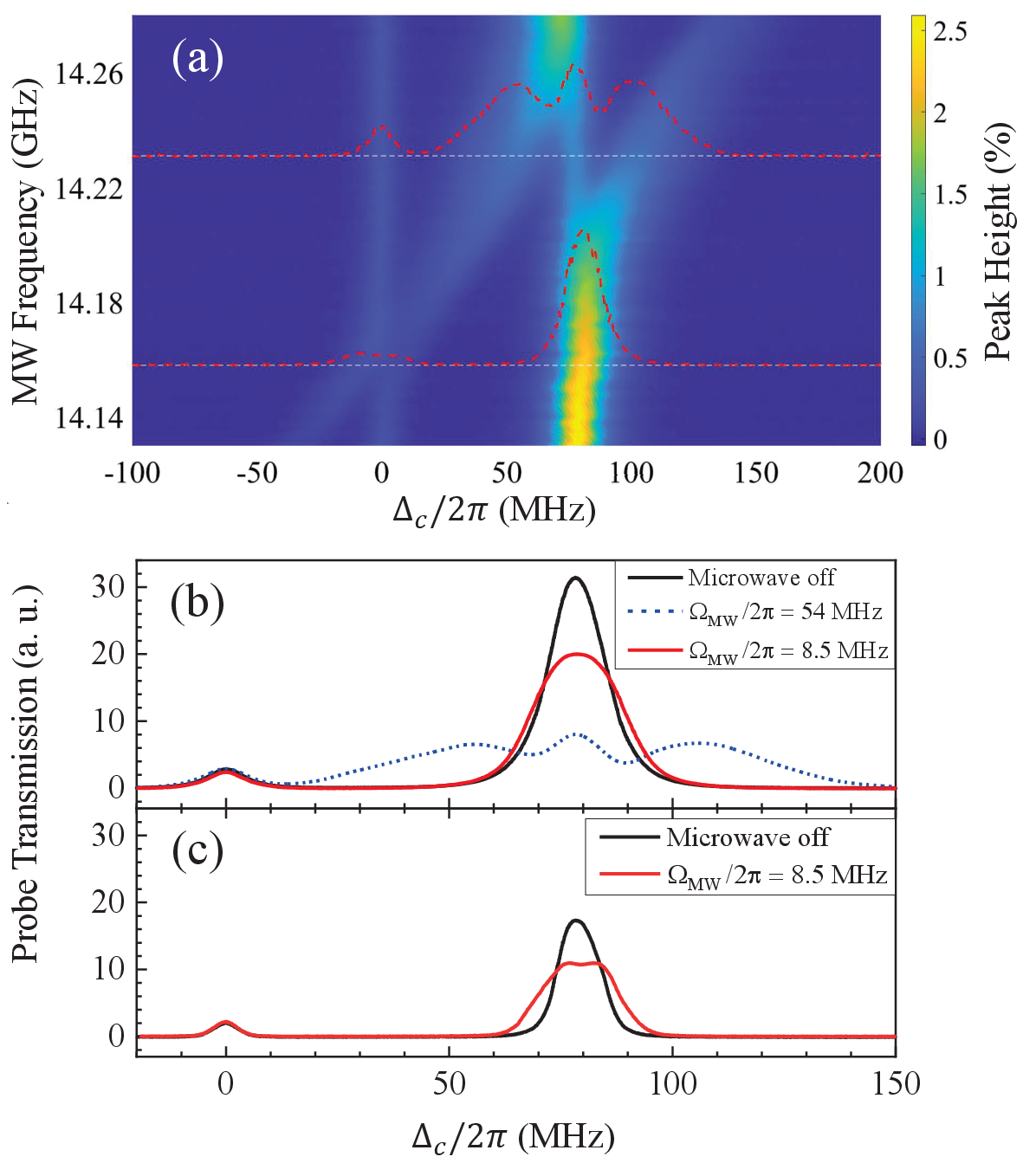}
\caption{Autler-Townes splitting of Rydberg-EIT spectra. (a) EIT transmission as a function of coupling laser detuning and MW frequency for the $|53D\rangle$ $\rightarrow$ $|54P_{3/2}\rangle$ transition. The panel also displays the spectral lines corresponding to the resonant MW frequencies of 14.231 GHz and 14.153 GHz. (b) EIT and AT splitting spectra with a fixed MW frequency of 14.231~GHz. The splitting of 54~MHz is observed, corresponding to an electric field strength of 12~mV/cm. The red solid line represents a field strength of 1.8~mV/cm ($\Omega_{MW}/2\pi=8.5$~MHz), where the splitting is unresolvable. (c) Spectra obtained with the probe laser intensity reduced by a factor of 5. This adjustment reduces power broadening of the EIT linewidth, allowing the 1.8~mV/cm field to be resolved. 
}
\label{fig:2}
\end{figure}
}

\newcommand{\Figbeat}{
\begin{figure}[t]
\centering
\includegraphics[width=1 \linewidth]{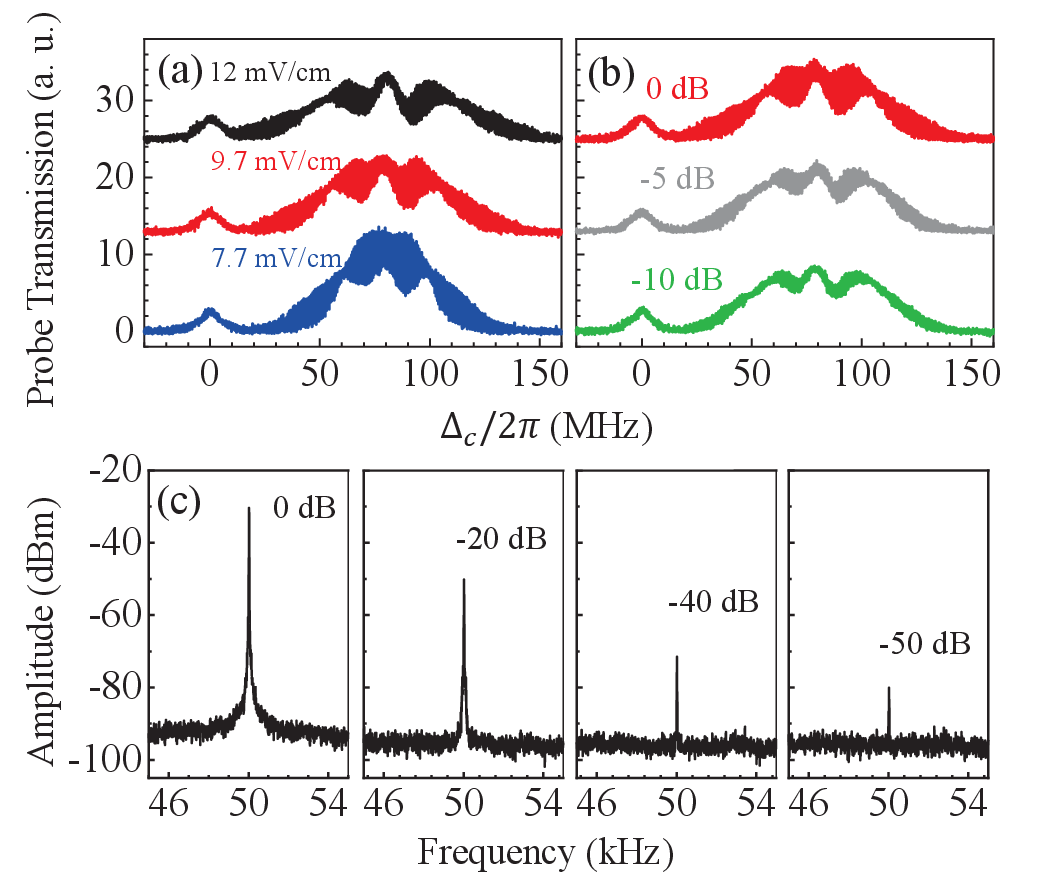}
\caption{Rydberg EIT spectra with beat signals measured under different field strengths. (a) Beat signal spectra as a function of LO field strength, with a fixed signal field strength of 1.9~mV/cm. (b) Beat signal spectra versus signal field strength, with a fixed LO field strength of 9.7~mV/cm. In these measurements, the beat frequency was set to $\Delta_{beat} = 7$~kHz to ensure clear resolution of the beat signals. The legend in (a) represents the LO field strength and that in (b) indicates the relative strength of the signal field, where 0~dB corresponds to the 1.9~mV/cm, determined from the AT splitting. (c) FFT of the probe transmission for different signal field strengths. These beat-note measurements were performed with fixed probe and coupling laser frequencies. The beat frequency was fixed at $\Delta_{beat}=50$~kHz. In the rightmost panel, the estimated signal field strength of 6.1~$\mu$V/cm yields a beat signal with an SNR of 16~dB.}
\label{fig:beat}
\end{figure}
}

\newcommand{\FigFour}{
\begin{figure}[t]
\centering
\includegraphics[width=1 \linewidth]{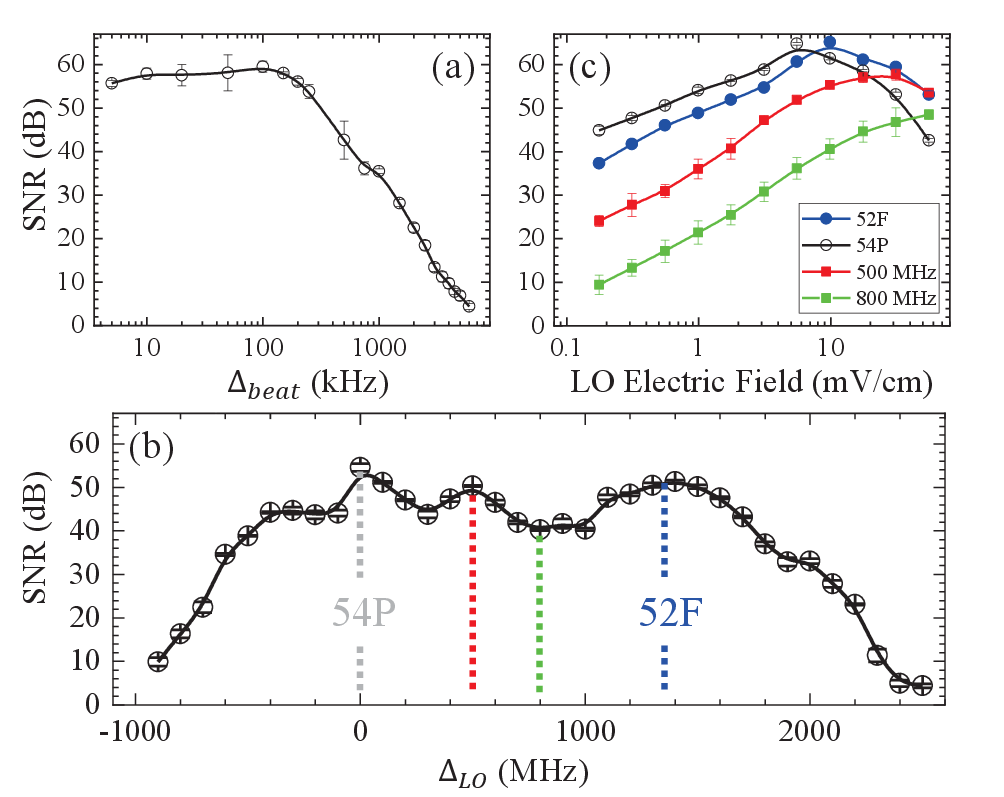}
\caption{(a) Beat-note SNR with varied beat frequency, $\Delta_{beat}$. The signal decay at higher $\Delta_{beat}$ indicates a maximum detectable frequency difference of approximately 6~MHz.
(b) The detectable range of varying the LO field frequency with $\Delta_{beat}=50$~kHz. Detuning 0 MHz was defined as the frequency of the transition from $|53D_{5/2}\rangle$ to $|54P_{3/2}\rangle$. Results show a detectable tuning range of 3 GHz (SNR $\geq 10$~dB) and a nearly constant high-sensitivity range of 2.2 GHz (SNR $\geq 40$~dB).
(c) Beat-note SNR as a function of LO field strength. The optimal LO field strength varies between the resonant AT transitions ($54P$ and $52F$) and the AC Stark transition (detuned by 500~MHz and 800~MHz from the $54P$ state).}
\label{fig:4}
\end{figure}
}

\newcommand{\FigFive}{
\begin{figure}[t]
\centering
\includegraphics[width=0.9 \linewidth]{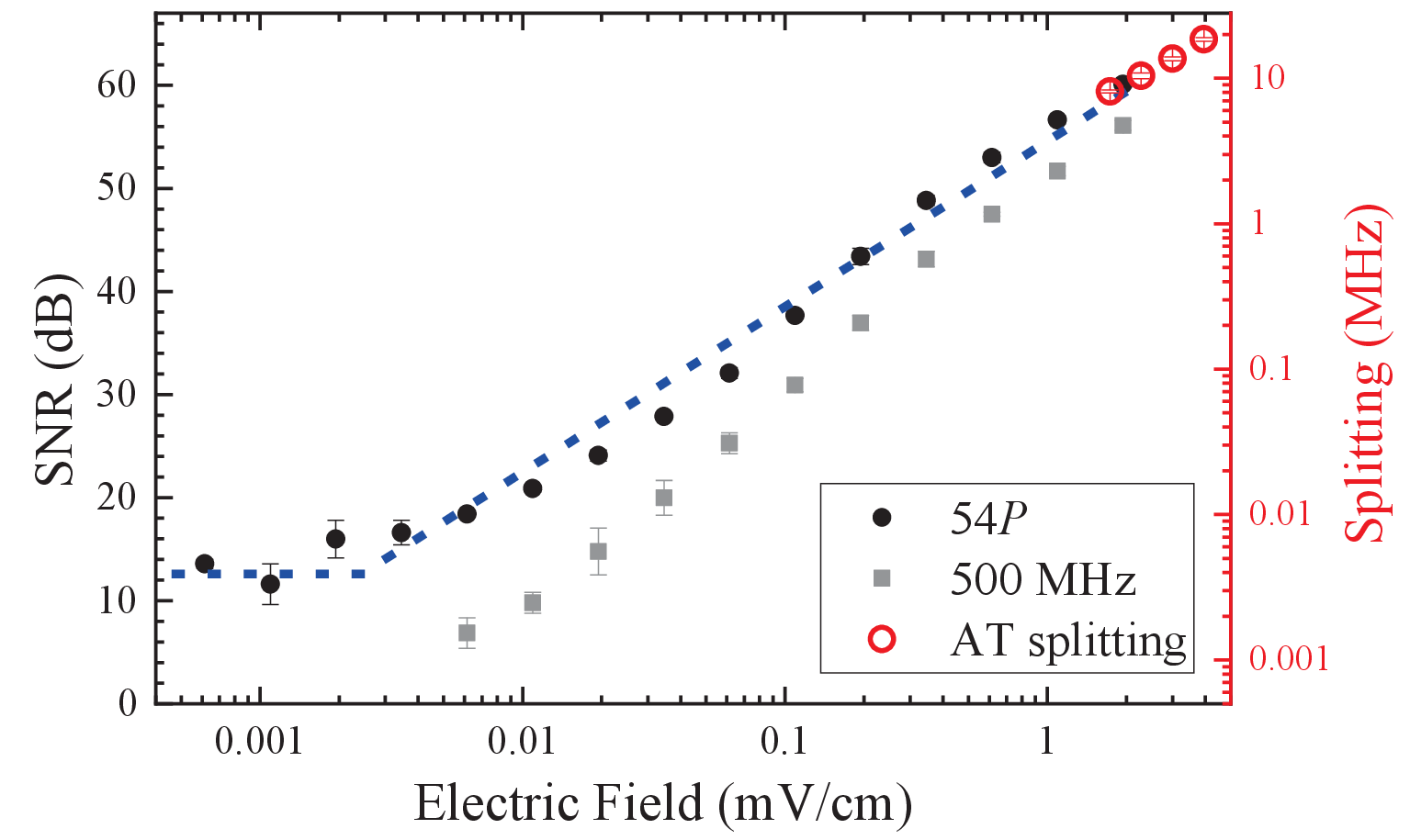}
\caption{Beat-note SNR as a function of signal electric field strength. Black circles correspond to the resonant AT transition ($54P$), while gray squares represent the AC Stark transition (detuned by 500~MHz from the $54P$ state). The electric field strength is derived from the measured AT splitting (right axis), indicated by the red solid circles.}
\label{fig:5}
\end{figure}
}
\begin{abstract}
We present a Rydberg atom-based microwave electric field sensor that achieves extended dynamic range and enhanced sensitivity across a broad bandwidth. By characterizing the Autler-Townes (AT) splitting induced by a single-tone microwave field, we demonstrate a spectroscopic method that simultaneously extracts both the microwave frequency and electric field strength directly from the splitting pattern.
We implement dual-tone heterodyne detection, achieving a minimum detectable field strength on the order of $\mu$V/cm and a sensitivity in the sub-$\mu\text{V/cm}/\sqrt{\text{Hz}}$ regime, while extending the operational bandwidth up to 3 GHz.
Through systematic characterization of frequency and power dependencies, we identify optimal operating conditions to minimize power broadening in the resonant AT regime and maximize sensitivity in the far-off-resonance AC Stark regime. The resulting platform combines high sensitivity, broad bandwidth, and a dynamic range of approximately 90~dB, establishing Rydberg atoms as practical sensors for precision electric field metrology.
\end{abstract}

\maketitle

\section{Introduction}
Recent advances in Rydberg atom-based detection have enabled highly accurate electromagnetic field sensing, offering significant advantages over traditional microwave receivers~\cite{Adams2020,Liu2023}. By utilizing the strong electric dipole moments of Rydberg atoms and the enhanced sensitivity provided by electromagnetically induced transparency (EIT), atomic sensors can precisely measure radio frequency (RF) fields across a wide frequency range from 100~MHz to over 1~THz~\cite{Jiao2017,Miller2016,Wade2017}. This technology is essential for data communication and remote sensing, drawing broad academic and industrial interest, and supports measurements of electric field amplitude, polarization~\cite{Sedlacek2013}, phase~\cite{Simons2019}, incident angle~\cite{Robinson2021, Schlossberger2025}, as well as microwave imaging~\cite{Holloway2014}. Recent work has demonstrated that microwave-assisted EIT spectroscopy enables high-precision measurement of Rydberg transition frequencies and quantum defects in atomic vapor cells~\cite{Li2021}. Furthermore, these advances have facilitated the demonstration of miniaturized quantum sensors, paving the way for compact and practical devices in real-world applications.

Key performance factors for electric field sensors include measuring frequency range, minimum detectable field, sensitivity, and the dynamic range of detectable power. For example, an electric field sensitivity of up to 1.5~$\mu$V/cm/$\sqrt{\rm{Hz}}$, 80-dB linear dynamic range, and over 1~GHz of continuous frequency range has been reported in Ref.~\cite{Liu2022}. To enhance the sensitivity of microwave electric field measurements with Rydberg EIT, several strategies have been developed. Selecting Rydberg states with higher principal quantum numbers increases the atomic dipole moment, thereby strengthening the interaction with the microwave field~\cite{Sedlacek2012}, while the choice of Rydberg angular states significantly affects the linearity and accuracy of microwave field measurements~\cite{Chopinaud2021}.
Furthermore, it has been demonstrated that optimizing the microwave frequency detuning can improve the sensitivity of electric field measurements based on EIT and Autler-Townes (AT) splitting~\cite{Simons2016}. Narrowing the EIT linewidth allows for more precise resolution of AT splitting, while optimizing beam alignment and polarization further improves EIT contrast and detection sensitivity. The measurement range can be extended to lower field strengths by employing dual-tone or auxiliary microwave driving schemes ~\cite{Jia2021,Jayaseelan2023}.
In addition, incorporating sub-wavelength resonator structures, such as split-ring resonators, enables the detection of much weaker fields~\cite{Holloway2022}. Multi-photon excitation schemes, including three-photon AT splitting in cold cesium Rydberg gases, offer enhanced capabilities for the control and precise detection of Rydberg states~\cite{Bai2022}. Advanced signal extraction techniques, such as lock-in amplification and superheterodyne detection, can isolate weak field-induced features from noise, achieving sensitivities down to the sub-$\mu$V/m/$\sqrt{\rm{Hz}}$ level~\cite{Jing2020,Ren2024,Liang2025}. Together, these methods enable highly accurate and sensitive microwave field measurements using Rydberg atom-based sensors. 
To extend the detectable frequency range, the Zeeman and AC Stark effects can be also employed to achieve continuous frequency tunability in measurements~\cite{Liu2022,Hu2022,Shi2023}. Reference \cite{simons2021continuous} demonstrated that introducing a second microwave field, resonant with an adjacent Rydberg transition, induces an AC Stark shift, enabling continuous and tunable frequency detection across neighboring Rydberg states. This improved tunability has opened up new opportunities for applications utilizing a planar microwave waveguide, which effectively operates across a frequency range of 0 to 20 GHz~\cite{Meyer2021}.

In this work, we develop a microwave electrometry technique using dual-tone heterodyne detection in a Rb vapor cell, building upon our prior Rydberg EIT studies~\cite{Su2022,Su2022_2,YuChi2025}. Our approach employs a strong local oscillator (LO) that interacts with a weak signal field detuned by a few tens of kHz, allowing the resulting beat notes to be detected via Rydberg-EIT spectroscopy.
For strong microwave fields, the signal strength is characterized by AT splitting, while in the weak-field regime,  the beat signals are analyzed using the Fast Fourier Transform (FFT) of the oscilloscope. Through systematic characterization, we identify optimal operating conditions by tuning the LO to resonant AT transitions and employing moderate field power to avoid power broadening. This optimization enables precise detection with a minimum detectable field of $2.4~\mu$V/cm and a sensitivity of $\rm{760~nV/cm/\sqrt{Hz}}$, while extending the dynamic range to over 60~dB and supporting MW frequency tuning up to 3~GHz. Furthermore, employing the AT splitting regime with a single microwave field extends the measurement capability, resulting in a total dynamic range of approximately 90~dB.
Ultimately, this platform combines high sensitivity, broad bandwidth, and wide dynamic range, positioning Rydberg atoms as practical sensors for spectrum monitoring, electromagnetic compatibility testing, and precision metrology.

\section{Setup and Methods}
\subsection{Energy level and laser system}
The experiment is carried out in a Rb vapor cell at around 60$^\circ$C. Two laser beams construct a three-level ladder-type EIT system, and microwave fields further couple two Rydberg states, as illustrated in Fig.~\ref{fig:setup}. The probe beam, with a wavelength of 780 nm (Toptica DL pro), was locked in the transition from the ground state $|5S_{1/2}, F=2 \rangle$ to the excited state $|5P_{3/2}, F'=3 \rangle$ via the saturation absorption spectroscopy. Meanwhile, the coupling beam, with a wavelength of 480~nm (Toptica DL pro HP~480), was frequency scanned across the transition from the excited state to the Rydberg states $|53D\rangle$. 
These two laser beams counter-propagate through a cylindrical Rb vapor cell with a length of 75~mm and a diameter of 25~mm. The laser polarizations are adjusted to $\sigma^+$ in order to optimize the Rydberg EIT signal~\cite{Su2022}. The probe beam has a $1/e^2$ waist radius of 810~$\mu$m and an intensity of 28~mW/cm$^2$, while the coupling laser has a radius of 820~$\mu$m and a power of 35~mW. A photodetector (PD3 in Fig.~\ref{fig:setup}(b)) monitors the probe transmission to detect the Rydberg EIT signal as the coupling laser is scanned, or the heterodyne beat-note signal when the coupling frequency is fixed. The beat-note signal is analyzed using the FFT function of an R$\&$S RTO64 oscilloscope. Laser frequency is calibrated using a separate Rydberg EIT reference system in an independent cylindrical cell, referred to as the Ref. Cell.

\FigOne

\subsection{Microwave system}
When Rydberg atoms are subjected to an AC electric field, their energy levels exhibit AT splitting. By analyzing the resulting spectral peaks, the strength of the applied electric field can be determined. In this study, we apply a microwave (MW) field to serve as the AC electric field and investigate this splitting phenomenon.
However, the sensitivity of a conventional AT-based system is fundamentally limited by the resolution of EIT spectral features, specifically the linewidth, which is constrained by the atomic decay rate.
To address the limited frequency detection range, we introduce two microwave fields to generate interferences in the AT spectra, thereby extending the detectable frequency range of electric field measurements. 
A strong field, referred to as the local oscillator (LO) field, shifts the Rydberg level into a high-sensitivity region. 
A weak signal field, detuned by several tens of kHz from the LO field, $\Delta_{beat}$, is mixed with the LO field within the Rydberg system to produce beat-note signals, which are then detected using Rydberg EIT spectroscopy.

The signal and LO fields are produced by two separate signal generators. The LO field is generated by an R$\&$S SGS100A signal generator, passed through a frequency multiplier (ZX90-2-24-S+), and subsequently amplified by an RF amplifier (ZVE-3W-183+).
The signal microwave is generated using a Rigol DSG830 signal generator, cascaded with two frequency multipliers (ZX90-3-812-S+ and ZX90-2-24-S+), and further amplified by an RF amplifier (ZX60-83LN-S+). Two RF attenuators (Vaunix LDA-5018V) are incorporated to finely adjust the output power levels. This configuration enables precise control over the distribution and intensity of the MW fields within the vapor cell, thereby ensuring accurate experimental measurements.
These two MW fields are combined using a power combiner and emitted from a single horn antenna positioned 50~mm from the vapor cell. The antenna's orientation is adjusted to optimize the MW polarization, thereby maximizing the spectroscopic signal.

These MW fields operate in the frequency range of 13.3 to 16.7 GHz, covering the resonant frequencies of the Rydberg transitions $|53D_{5/2}\rangle \rightarrow |54P_{3/2}\rangle$ (14.23 GHz) and $|53D_{5/2}\rangle \rightarrow |52F\rangle$ (15.59 GHz). The radial matrix elements for these transitions are calculated \cite{SIBALIC2017} as 3622~$ea_0$ and 3594~$ea_0$, where $e$ is the charge of an electron, $a_0$ is the Bohr radius. $\Delta_{LO}$ denotes the MW detuning to the Rydberg state $|54P_{3/2}\rangle$ or $|52F\rangle$. 
In the strong-field regime, the field strength can be directly derived from the measured AT splitting and the known radial matrix elements.
When the LO frequency is far-detuned from the target quantum state, the dominant interaction mechanism converts from resonant AT splitting to dispersive AC Stark shifting. 
In the weak-field regime where AT splitting is not resolvable, we employ a heterodyne detection technique. The resulting interference manifests as an oscillation in the atomic response, the amplitude of which allows for the precise determination of the weak electric field strength.

\subsection{Methods}
We employ Rydberg atom-based electrometry to measure microwave electric fields across a broad frequency range, exploiting both the AT splitting in the near-resonant regime and the AC Stark shift in the far-off-resonant regime. Using standard EIT with AT splitting techniques, the electric field strength can be measured from the observed splitting~\cite{Sedlacek2012},
\begin{equation}
\delta=\frac{\Omega_{MW}}{2\pi}=\frac{d|E|}{2\pi\hbar},
\label{eq:ATS}
\end{equation}
where $\Omega_{MW}$ is the Rabi frequency of the applied microwave field, $d$ is the atomic dipole moment, and $\hbar$ is the reduced Planck constant. The measurement resolution of the AT splitting is limited by the linewidth and contrast of the Rydberg EIT signal, which can be optimized by adjusting the laser power and atomic density~\cite{Su2022,Su2022_2}.

For heterodyne detection with two microwave fields near resonance with the Rydberg transition, an LO field at frequency $\omega_{LO}$ and a signal field at $\omega_{Sig}$ are simultaneously applied. The total electric field is given by
\begin{equation}
E_{tot}(t) = E_{LO}\cos(\omega_{LO}t + \phi_{LO}) + E_{Sig}\cos(\omega_{Sig}t + \phi_{Sig}),
\label{eq:totalE}
\end{equation}
where $E_{LO}$ and $E_{Sig}$ are the amplitudes, and $\phi_{LO}$ and $\phi_{Sig}$ are the phases of the LO and signal fields, respectively. For a strong LO field ($E_{LO} \gg E_{Sig}$) with a beat frequency $\Delta\omega = \omega_{Sig} - \omega_{LO}$, the instantaneous field amplitude can be approximated as
\begin{equation}
|E_{tot}(t)| \approx E_{LO} + E_{Sig}\cos(\Delta\omega t + \Delta\phi),
\label{eq:beatfield}
\end{equation}
where $\Delta\phi = \phi_{Sig} - \phi_{LO}$ is the phase difference. In this resonant regime, the AT splitting is proportional to the field amplitude $|E_{tot}(t)|$. The strong LO field biases the atom to a specific splitting width, while the weak signal field modulates this splitting at the beat frequency $\Delta\omega$ with an amplitude proportional to $E_{Sig}$. This allows for linear detection of the signal field amplitude.

For fields far detuned from the Rydberg transition, the atomic response is governed by the AC Stark shift, which depends on the square of the total electric field,
\begin{equation}
\delta_{Stark}(t) = -\frac{1}{2}\alpha E_{tot}^2(t),
\label{eq:starkshift}
\end{equation}
where $\alpha$ is the polarizability of the Rydberg state. After time-averaging over the fast carrier oscillations ($\sim\omega_{LO}$), the squared field retains the slowly-varying beat envelope, 
\begin{equation}
E_{tot}^2(t) \approx \frac{E_{LO}^2}{2} + \frac{E_{Sig}^2}{2} + E_{LO}E_{Sig}\cos(\Delta\omega t + \Delta\phi).
\label{eq:totalEsquared}
\end{equation}
In this dispersive regime, the atom exhibits a non-linear quadratic response to the total electric field. The Stark shift modulates the EIT transmission at the beat frequency $\Delta\omega$. Crucially, the beat-note amplitude in Eq.~(\ref{eq:totalEsquared}) is proportional to the product $E_{LO}E_{Sig}$. This demonstrates that the strong LO field amplifies the weak signal contribution by a factor of $E_{LO}/E_{Sig}$ (relative to direct detection of $E_{Sig}^2$), providing the heterodyne gain necessary for high-sensitivity detection.

In both regimes, the heterodyne detection mechanism relies on the same principle: the beat note between the LO and signal fields produces a time-varying atomic response at the angular beat frequency $\Delta\omega$. This corresponds to a measurable modulation frequency $\Delta_{beat} = \Delta\omega / 2\pi$. Since $\Delta_{beat}$ (typically tens of kHz) is orders of magnitude lower than the EIT linewidth ($\Gamma_{EIT} \sim$ MHz), the atoms follow the instantaneous field envelope adiabatically. Consequently, the beat-note signal can be directly retrieved from the probe transmission using a lock-in amplifier or spectrum analyzer. This enables phase-sensitive heterodyne detection, where both the amplitude and phase of the signal field are preserved. By operating across both the near-resonant AT splitting regime and the far-detuned AC Stark shift regime, our method achieves continuous frequency coverage up to 3 GHz. This dual-regime approach combines the high sensitivity of resonant detection with the broad bandwidth of off-resonant operation, providing a versatile platform for wideband RF field sensing.

\section{Results}
\FigTwo
\subsection{Autler-Townes Splitting Analysis}
To establish the relationship between applied MW power and electric field strength, we performed calibration measurements using AT splitting. 
Figure~\ref{fig:2} displays the transmission spectra obtained by scanning the coupling laser frequency across the Rydberg transitions. The coupling detuning, $\Delta_c/2\pi$, is calibrated using the known $78.3$~MHz fine-structure splitting between the $|53D_{3/2}\rangle$ and $|53D_{5/2}\rangle$ states.
The MW frequency is scanned across the transition from $|53D\rangle$ to $|54P_{3/2}\rangle$. Resonances are observed at $14.231$~GHz and $14.153$~GHz, corresponding to the $|53D_{5/2}\rangle \rightarrow |54P_{3/2}\rangle$ and $|53D_{3/2}\rangle \rightarrow |54P_{3/2}\rangle$ transitions, respectively. The figure also shows the spectral lines associated with these resonant MW frequencies. The $78$~MHz difference between these resonances is consistent with the fine-structure splitting of the $|53D\rangle$ state. The estimated power generated by the horn antenna in Fig.~\ref{fig:2}(a) is approximately $15$~dBm.

To accurately determine the electric field strength associated with this power, we investigate the power-dependent AT splitting at the resonant frequency. Figure~\ref{fig:2}(b) presents the Rydberg EIT-AT spectra measured at various MW powers, with the MW frequency fixed at 14.231 GHz. We determined the effective Rabi frequencies and electric field strengths from the measured splitting using Eq.~(\ref{eq:ATS}).
For example, the blue-dashed trace shows an AT splitting of 54 MHz, corresponding to a Rabi frequency $\Omega_{MW}/2\pi=54$~MHz. From the $|53D_{5/2}\rangle \rightarrow |54P_{3/2}\rangle$ radial matrix element and an assumed unity Clebsch-Gordan coefficient, we obtain a field strength of 12~mV/cm. Reducing the microwave power by 16 dB yields a field of 1.8 mV/cm, where the splitting becomes unresolvable (red solid trace). 
Because the horn antenna is tilted or rotated, the MW field vector acquires multiple components relative to the laser polarization, complicating the atomic transition. The simultaneous excitation of both EIT and AT pathways, caused by these mixed vector components, leads to a superposition of resonances in the observed spectrum. An unexpected peak, attributed to the unperturbed EIT pathway, indicates an imperfect polarization match between the optical and microwave fields.
A similar phenomenon has been discussed in Ref.~\cite{Sedlacek2013}. 

The power-dependent measurements establish a direct relationship between AT splitting and field strength, enabling quantitative measurements.
The resolution of the microwave electric field strength is inherently limited by the EIT linewidth. Initially, with a measured linewidth (full-width at half-maximum) of 13 MHz in the absence of the MW field (shown as the black line in Fig.~\ref{fig:2}(b)), the minimum detectable field corresponds to approximately $\Omega_{MW}/2\pi=$ 13~MHz. To improve sensitivity, we reduced the probe laser intensity by a factor of 5, thereby suppressing power broadening~\cite{Wu2017, Su2022}. Under these optimized conditions, the minimum detectable microwave Rabi frequency improves to approximately $\Omega_{MW}/2\pi = 8.5$~MHz. This corresponds to an electric field of 1.8~mV/cm, which is the same power level depicted by the red line in Fig.~\ref{fig:2}(b). Notably, reducing the probe power may not improve performance, as it excessively degrades the EIT peak. Consequently, for the subsequent detection of dual-tone heterodyne, we retained the original probe power (before the fivefold reduction) to maximize the beat signals. While optimizing the EIT laser power is critical, a comprehensive investigation is beyond the scope of this study and remains a subject for future work.

\subsection{Dual-Tone Heterodyne Detection}
While the AT splitting technique enables microwave electric field measurements down to several mV/cm, the detectable frequency range is fundamentally limited by the Rydberg EIT linewidth. To overcome this limitation, we employ a dual-tone microwave scheme consisting of a strong local oscillator (LO) field and a weak signal field. This heterodyne approach exploits the interference between the two microwave fields, generating beat signals described by Eq.~(\ref{eq:beatfield}) in the near-resonant AT regime and Eq.~(\ref{eq:totalEsquared}) in the far-off-resonant AC Stark shift regime. These beat signals are detected via the Rydberg EIT, offering enhanced sensitivity compared to conventional AT splitting measurements. Figures~\ref{fig:beat}(a) and \ref{fig:beat}(b) display clear beat oscillations at $\Delta_{beat}=7$~kHz, generated by the interference between the LO and signal fields at $\sim$14.231 GHz . The spectrum scan rate was set to a few Hz, ensuring it was sufficiently slow compared to the beat frequency to resolve the oscillations accurately. 

To characterize the dual-tone response, we systematically varied the LO and signal field powers, as shown in Fig.~\ref{fig:beat}. Figure~\ref{fig:beat}(a) illustrates the Rydberg EIT spectra with beat signals as a function of LO field strength (12, 9.7, 7.7~mV/cm from top to bottom) at a fixed signal field of 1.9~mV/cm. Conversely, Fig.~\ref{fig:beat}(b) displays the response as a function of signal field strength (1.9, 1.1, 0.61~mV/cm) with the LO field fixed at 9.7~mV/cm. As shown in Fig.~\ref{fig:beat}(a), the LO field strength is sufficiently strong to induce AT splitting. In contrast, the signal field is kept weak enough to avoid perturbing this splitting, as demonstrated in Fig.~\ref{fig:beat}(b). Consequently, the signal field serves only to enhance the beat signal amplitude without altering the underlying dressed-state structure. All field strengths in these experiments were calibrated using single-tone microwave AT splitting.

\Figbeat

For quantitative analysis, the coupling laser was tuned to the AT resonance, which corresponds to the maximum beat signal. The beat signal was then monitored using the fast Fourier transform (FFT) function of the oscilloscope (R$\&$S RTO64). Figure~\ref{fig:beat}(c) displays the FFT of the detected probe transmission, centered at $\Delta_{beat}=50$~kHz. In the rightmost panel, corresponding to -50 dB (signal field strength of 6.1~$\mu$V/cm), the beat signal exhibits a signal-to-noise ratio (SNR) of 16 dB.
Our dual-tone configuration enables precise detection of weak fields without perturbing the atomic energy levels.
Building on this capability, the next subsection demonstrates how this heterodyne approach can be extended to achieve continuously tunable detection across a broad microwave spectrum.

\subsection{Frequency and Dynamic Range Extension}
To evaluate the detection bandwidth of the dual-tone system, we characterized the beat signal response as a function of the frequency detuning between the two microwave fields. Figure~\ref{fig:4}(a) displays the detectable signal frequency range for dual-tone heterodyne detection. In these measurements, the LO field frequency was fixed at the Rydberg transition resonance ($|53D_{5/2}\rangle \rightarrow |54P_{3/2}\rangle$, 14.231 GHz), while the signal field frequency was varied to produce beat frequencies $\Delta_{beat}$ ranging from 5~kHz to 6~MHz. The strengths of the LO and signal MW fields were maintained at 9.7 and 1.9~mV/cm, respectively. The results confirm that the Rydberg EIT-based heterodyne detection scheme maintains a robust response over a frequency range. We observed a nearly constant SNR for beat frequencies below 100~kHz. The measurements were conducted down to 5~kHz, which does not represent a fundamental lower detection limit. Based on the observed decay trend, we estimate the upper detection bandwidth to be approximately 6~MHz, beyond which the beat signal approaches the noise floor. This upper limit is governed by the transient response time of the EIT system, which is on the order of a few $\mu$s~\cite{YuChi2025}. This response time is determined by the atomic coherence decay rate, arising from intrinsic decoherence processes such as atomic collisions and natural linewidths.
\FigFour

The interference signal depends on the beat frequency, whereas the absolute frequency of the LO field determines whether the Rydberg system exhibits resonant AT splitting or detuned AC Stark shifts. We verified this frequency response by sweeping the LO field frequency while maintaining a constant beat frequency $\Delta_{beat}=50$~kHz, as shown in Fig.~\ref{fig:4}(b). 
Zero detuning (0 MHz) was defined at the $|53D_{5/2}\rangle$ to $|54P_{3/2}\rangle$ transition frequency (14.23 GHz). When the detune is set to +1360 MHz, the LO field frequency approaches the $|53D_{5/2}\rangle$ to $|52F\rangle$ transition at 15.59 GHz. 
The results demonstrate a maximum detectable frequency tuning range of up to 3 GHz with SNR above 10~dB, and a high-sensitivity operating range of approximately 2.2~GHz where the signal power exceeds 40~dB. We observed symmetric signal decay at approximately $\pm800$~MHz detuning from each transition resonance: red-detuned from $|54P_{3/2}\rangle$ and blue-detuned from $|52F\rangle$. The LO field frequency was positioned between the two states, allowing both transitions to contribute to the signal in the detuned regime. This approach preserved sensitivity and prevented continuous signal decay.

Furthermore, the performance of heterodyne Rydberg atom sensors relies on optimizing the LO field power to balance sufficient coupling strength for AC Stark shift detection. To determine the optimal operating conditions, we systematically characterized the signal response as a function of the LO field strength in different frequency regimes, as shown in Fig.~\ref{fig:4}(c). 
The microwave fields were set to four different frequencies: resonant with the $|54P_{3/2}\rangle$ and $|52F\rangle$ transitions, and detuned by 500~MHz and 800~MHz from $|54P_{3/2}\rangle$, while maintaining a constant beat frequency of 50~kHz.
Both the resonant AT transitions (black and blue circles) and the detuned AC Stark regime (red and green circles) exhibit an increase in signal strength with increasing LO field power. However, they differ significantly in their optimal power requirements. The resonant AT transitions achieve optimal signal strength at lower LO powers, as their performance is eventually limited by power broadening ($\propto\Omega_{MW}$), which degrades signal contrast and spectral resolution. In contrast, the detuned AC Stark regime operates in the far-detuned limit where the AC Stark shift scales as $\propto E^2/\Delta_{LO}$ without significant power broadening. Consequently, the signal strength in this regime continues to improve with higher LO power as estimated in Eq.~(\ref{eq:totalEsquared}). As the field-induced energy shift grows, the beat-note SNR reaches an optimized power level that is 15 dB stronger than that of the resonant AT regime. Thus, the resonant regime is broadening-limited, whereas the detuned regime is power-enhanced.

\FigFive

Using this dual-tone heterodyne detection scheme, we measured the beat-note SNR as a function of signal electric field strength, as shown in Fig.~\ref{fig:5}. Black circles correspond to the resonant AT transition ($54P$) with an optimized LO field strength of 5.5~mV/cm, while gray squares represent the AC Stark transition (detuned by 500 MHz from the $54P$ state) with an optimized LO field strength of 17~mV/cm. The electric field strength (right axis) was derived from the measured AT splitting, indicated by the red circles. We observed a minimum detectable field of 2.4~$\mu$V/cm for the resonant AT transition, corresponding to a sensitivity of 760~nV/cm/$\sqrt{\text{Hz}}$ at a 10 Hz resolution bandwidth. In the off-resonant AC Stark regime, the minimum detectable field was 6.1~$\mu$V/cm, with a corresponding sensitivity of 1.9~$\mu$V/cm/$\sqrt{\text{Hz}}$. Based on these measurements, the current setup yields a dynamic range of approximately 60~dB. By utilizing the AT splitting regime with a single microwave field at the maximum field strength of 55~mV/cm, the dynamic range can be further extended to approximately 90~dB. This combination of high sensitivity and extended dynamic range positions the dual-tone heterodyne method as a versatile tool for precision electrometry across a wide spectrum of field strengths

\section {Conclusions}\label{conclusions}
This work presents a comprehensive investigation of Rydberg atom-based microwave electric field sensing, significantly expanding the capabilities of atomic receivers. We characterized the AT splitting induced by a single microwave field, establishing a self-calibrating spectroscopic method to simultaneously determine both resonant frequency and electric field strength. To further enhance sensitivity, we implemented a dual-tone heterodyne detection scheme. Through systematic optimization of the local oscillator frequency, beat frequency, and field power, this scheme yields a minimum detectable field of 2.4~$\mu$V/cm and the sensitivity of $\rm{760~nV/cm/\sqrt{Hz}}$. By integrating standard AT splitting with heterodyne detection, our system achieves a dynamic detection range of up to 90 dB and a wide tuning range of 3 GHz (SNR $\geq$ 10 dB). These techniques collectively address the performance requirements of practical applications, including spectrum monitoring, electromagnetic compatibility testing, and precision metrology.
\begin{acknowledgments}
\label{acknowledgements}
This work was supported by Grants Nos. 113-2112-M-110-010 and 114-2112-M-110-005 of the National Science and Technology Council, Taiwan. 
\end{acknowledgments}


\bibliography{MWreferences}

\end{document}